\RequirePackage{fix-cm}

\documentclass[smallextended,natbib]{svjour3}

\usepackage{amsmath}
\usepackage{amssymb}
\usepackage{algorithm2e}
 
\usepackage{graphicx}
\usepackage{color}
\usepackage{url}
\usepackage[]{geometry}

\newcommand{\doi}[1]{(doi:#1)}

\newcommand{\rev}[1]{\textcolor{black}{#1}}

\newcommand{\bz}{\mathbf{z}}

\title{Importance sampling for partially observed temporal epidemic models}

\author{Andrew J. Black} 

\institute{A. J. Black \at
              School of Mathematical Sciences, 
              University of Adelaide,
              Adelaide, SA 5005,
              Australia.\\
              ACEMS, School of Mathematical Sciences, University of Adelaide,
              Adelaide, SA 5005,
              Australia.\\
              \email{andrew.black@adelaide.edu.au}           %  \\
}

\date{\today}

\begin{document}

\maketitle
 
\begin{abstract}

We present an importance sampling algorithm that can produce realisations of Markovian epidemic models that exactly match observations, taken to be the number of a single event type over a period of time. 
The importance sampling can be used to construct an efficient particle filter that targets the states of a system and hence estimate the likelihood to perform Bayesian inference.
When used in a particle marginal Metropolis Hastings scheme, 
the importance sampling provides a large speed-up in terms of the effective sample size per unit of computational time, compared to simple bootstrap sampling.
The algorithm is general, with minimal restrictions, and we show how it can be applied to any continuous-time Markov chain where we wish to exactly match the number of a single event type over a period of time.
\end{abstract}

\section{Introduction}

Many epidemic models are most naturally described by continuous-time Markov chains \citep{Keeling:2007,BM12}. These capture the discrete nature of the individuals, which is important when considering smaller populations, as well as the random nature of the underlying events. 
Bayesian inference using these models is difficult because, apart from when the state space is small \citep{BR13,Black:2017}, the transition density for such models cannot be evaluated point-wise. Thus many modern methods for performing inference using these models rely on simulating from the underlying model---sampling the transition density instead of evaluating it---which is typically quite simple \citep{Golightly:2011}.

One such method is particle marginal Metropolis Hastings (pmMH) \citep{Andrieu:2010}. This can be understood as a basic Metropolis-Hastings algorithm targeting the parameter posterior of the model, but where the likelihood is replaced by an unbiased estimate from a particle filter, which is a form of sequential Monte Carlo (SMC) \citep{Doucet2001,Doucet2009}. The popularity of pmMH stems from it targeting the exact posterior distribution for the parameters of the model, despite the estimate.
The challenge in implementing this method is that the mixing of the chain depends  strongly on the variance of the likelihood estimate \citep{Doucet2015,Sherlock2015}, which in turn depends on the performance of the particle filter.

The simulations of the discrete-state model are normally done using the stochastic simulation algorithm (SSA)\footnote{Also known as the Gillespie algorithm or the Doob-Gillespie algorithm.} \citep{Gillespie:1976}, which is an example of a bootstrap filter \citep{Gordon1993}. The problem with this approach is that if the observed events are rare, or the state space is large, the number of particles needed to estimate the state and hence the marginal likelihood in any one step of the SMC becomes prohibitively large. This is a well known problem in SMC where the more accurate the observations, the worse the filter performs. So in epidemic models where we observe a component of the state exactly---for example the number of infection or recovery events over an interval of time (but not the exact times at which they occur)---the cost of producing simulations that match this data becomes high. 
This can be mitigated to some extent by assuming or adding noise on top of the observations, essentially increasing the likelihood of a particle matching \citep{Golightly:2011}. Whether this is reasonable, or not, is a modelling decision, but will inevitably add extra variance to the parameter estimates. A better approach is to use importance sampling to generate realisations of the process that exactly match the observations. Importance sampling works by changing the rules by which a process evolves so as to make a rare event more probable \citep{Kroese:handbook}. This bias is then  corrected for in the calculation of the likelihood.

In this paper we present a simulation algorithm that implements importance sampling to produce realisations of complex stochastic epidemic models that match observations exactly. This builds substantially on the earlier work of \citet{McKinley:2014}. By modifying and extending the basic ideas of \citeauthor{McKinley:2014}, the resulting algorithm can be easily applied to quite complex models and does not suffer the numerical instabilities that are inherent to the original algorithm. 

We begin by describing the basic importance sampling idea for the simplest one-dimensional model. We then show how this can be generalised to a number of more complex multi-dimensional epidemic models, where our data are the observations of one component of the state, typically the number of a certain transition or event. 
%These algorithms produces realisations that match all observations exactly. 
Finally, we illustrate the use of importance sampling in a particle filter to perform inference on a number of outbreak time series using a model that accounts for different infectious phases. The resulting posterior distributions are compared to results from using an alive particle filter \citep{DelMoral:2015}, which uses the SSA for sampling. \rev{The version using importance sampling achieves a large speed-up in the effective sample size per unit of computation time using the same number of particles, and this speed-up increases as the size of the data set grows.} MATLAB and C code is provided for all of these methods as part of the \citet{Epistruct} project.

\section{Importance sampling}
\label{sec:decay}

We first introduce the basic importance sampling idea for a continuous-time Markov chain using a very simple example. 
Consider a simple model of a decay process where there is a collection of $M$ objects initially, which each decay independently at rate $\gamma$. We define $Z(t)$ to be the number of decay events by time $t$, hence the number of objects left is $X(t) = M-Z(t)$. The rate of an event is therefore
\begin{equation}
	a(Z(t)) = \gamma (M-Z(t)).
	\label{eq:decay_rate}
\end{equation}
Assume that over some interval of time, which without loss of generality we take as $[0,1]$, we observe $y$ events. We then wish to calculate a Monte Carlo estimate of the likelihood of this observation\footnote{In fact this can be done analytically for this model.}, which corresponds to one step in a sequential Monte Carlo routine \citep{Doucet2001,Doucet2009}. The likelihood of our observation can be written
\begin{equation}
	p(y) = p(y|z_1) p(z_1 | z_0),
	\nonumber
\end{equation}
where $z_j = Z(j)$ is the state of the system at time $t=j$.
As we observe the state of the system exactly, the observation density is
$$
p(y | z_1 )
= 
\delta_{z_1,\,y}
=
\begin{cases}
1  \quad \text{if $z_1$ matches $y$}, \\
0 \quad \text{otherwise}.
\end{cases} 
$$
Given the initial state of the process, we can sample from the transition density by using the SSA ($z_t^{(i)} \sim p(z_t | z_0)$), where $z_t^{(i)}$ is the state of the $i$th realisation, or particle, at time $t$ \citep{Gillespie:1976,Golightly:2011}. A Monte Carlo estimate of the likelihood is simply \citep{Kroese:handbook},
 \begin{equation}
	\hat{p}(y) = \frac{1}{N} \sum_{i=1}^N \delta_{z_1^{(i)}, y},
	\nonumber
\end{equation}
where $N$ is the total number of realisations produced.

\begin{figure}[ht!]
	\centering
	\includegraphics[width=0.5\textwidth]{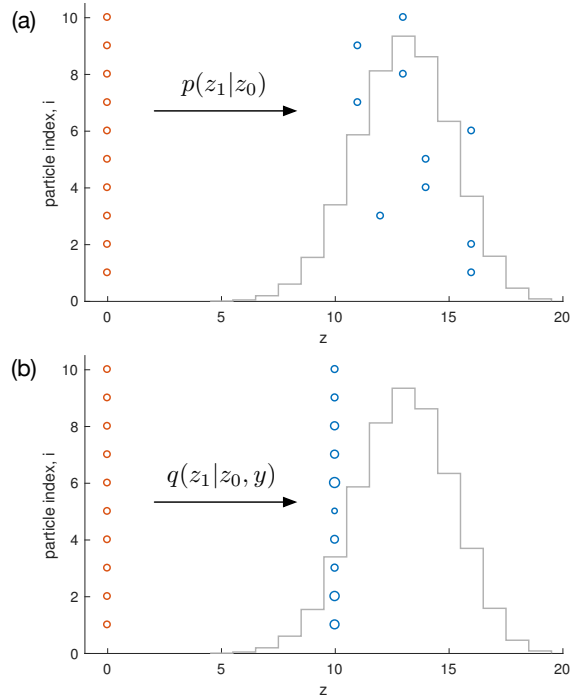}
	\caption{Illustration of bootstrap sampling from the transition density (a) and importance sampling (b), with $N=10$, $y=10$ and $\gamma=1$. The red circles indicate the particles initially; the blue circles show the particles after propagation by 1 day. The grey plot shows the true transition density. The importance sampling makes all the particles end in the observed state, but the particles have different weights (indicated by the size of the circles). Note that in panel (a) the estimate of the likelihood is 0 as none of the particles match the observation, and hence all particles would be assigned weight 0.}
	\label{fig:decay}
\end{figure}

As there is only one type of event, simulation of a realisation of the process is straightforward. Starting at $t=0$, the time to the next decay event is exponentially distributed with rate parameter $a$, given in Eq.~\eqref{eq:decay_rate},
\begin{equation}
	t' \sim \text{Exp}(a).
	\nonumber
\end{equation}
Hence we generate times and increment the variables $Z \leftarrow Z+1$ and $t\leftarrow t+t'$, and keep repeating this, until the next generated time is greater than the observation window ($t+t' > 1$), then the algorithm stops. A realisation of the process can then be specified by the initial state and the set of times, $\{t_1,\dots,t_n \}$, at which events occur.
Sampling from the transition density is illustrated in Figure \ref{fig:decay}(a), using $N=10$ realisations. If the observation was, for example, $y=10$ then none of the particles match this, hence their weight would be zero and the estimate of the likelihood zero.

% Only a single particle ends in the observed state ($y=10$), so our estimate of the likelihood is $1/10$. The transition density, $p(z_1|z_0)$, is also plotted for comparison, we can see that most realisations will not match our observation; this could be either because the observation is rare, or the parameter is not conducive to generating the observation.

% \rev{Could also add the bit about adding noise to improve the performance at the expense of increased variance.}
%This also illustrates the more general point that as the size of the state space expands the particles cover it less well.

Instead we can use importance sampling to estimate the likelihood \citep{Kroese:handbook}. This means we sample a realisation of the process according to an importance sampling distribution, $z_1^{(i)} \sim q(z_1|z_0,y)$, but this has to be taken into account in our estimate to recover the correct likelihood:
\begin{equation}
		\hat{p}(y) = \frac{1}{N} \sum_{i=1}^N  p(y |z^{(i)}_1) \frac{p(z_1^{(i)} | z_0) }{q(z_1^{(i)} |z_0,y)}.
		\label{eq:decay_weight}
\end{equation}
The key insight developed by \citet{McKinley:2014} is that it is possible to design simulation algorithms such that the observation likelihood $p(y | z^{(i)}_1)=1$, i.e., all the realisations end in a state consistent with the observation. Eq.~\eqref{eq:decay_weight} then reduces to 
\begin{equation}
		\hat{p}(y) = \frac{1}{N} \sum_{i=1}^N \frac{p(z_1^{(i)} | z_0) }{q(z_1^{(i)} |z_0,y)} 
		= 
		\frac{1}{N} \sum_{i=1}^N
		w_i,
		\label{eq:decay_weight2}
\end{equation}
where the ratio of the transition density, $p$, to the importance sampling density, $q$, is known as the weight, and can be calculated iteratively as the simulation progresses.

One choice of importance sampling density for this decay model is particularly simple. If $y$ events are observed then we first generate the times of the events distributed uniformly over the observation interval, and these are sorted,
\begin{equation}
	\tau = \{t_1,\dots,t_y\}.
\end{equation}
These are known as order statistics \citep{David2005}  and can be generated by sorting uniform random numbers or by employing a dedicated algorithm \citep{Kroese:handbook}. 
With the times, $\tau$, generated, all that remains is to calculate the weight of the realisation. The importance density of any particular realisation is just 
\begin{equation}
	q(\mathbf{\tau}) = y!,
	\label{eq:init_import}
\end{equation}
which follows from considering the joint distribution for the $y$ uniform random variables. 
Under the original process---i.e., the one where events occur at a rate given by Eq.~\eqref{eq:decay_rate}---the time to the next event is distributed exponentially, with pdf
\begin{equation}
	f(t) = a e^{-a t}.
	\nonumber
\end{equation}
Thus for a given set of times, $\tau$, the transition density under the original process is
\begin{equation}
	p(\tau)
	 =
	  e^{-a(Z(t_y))(1-t_y)}
	 \prod_{i=1}^{y} a(Z(t_{i-1})) e^{-a(Z(t_{i-1})) (t_i-t_{i-1}) },
\end{equation}
where $t_0=0$ and the term at the front of the expression is the probability of no further events in the interval $[t_y,1]$.
%\rev{Can also write this in terms of the intervals or dwell times. I think we still need to define the last one of these carefully as interval of time until the end of the observation window.}
The recursive nature of this expression means that it is simple to evaluate this iteratively---this is seen clearly in the code provided---and this is exploited in later simulation algorithms. In practice we work with the log of $q$ and $p$ to avoid numerical issues. 
Figure \ref{fig:decay}(b) shows 10 realisations produced using importance sampling simulations for the same example as before. By construction, all particles end in the observed final state, but have different weights. 
% If this was being run as part of an SMC scheme, then typically now we would re-sample these particles, in fact for this model we can just replace then, as we know the state exactly. 

% \begin{figure}[th]
% 	\centering
% 	\includegraphics[width=0.66\textwidth]{figs/figure2.pdf}
% 	\caption{Show weights from importance sampling scheme and how particles now all end up in the correct state. Could combine with the first figure in a column if that looked better????. The marginal likelihood is estimated as ....}
% 	\label{fig:is1}
% \end{figure}

%\section{Exact sampling for epidemic models}

\section{Application to epidemic models}

\subsection{SIR model}
\label{sec:SIR_model}

We now take the basic idea developed in the previous section---that we can simulate realisations that match the observations exactly---and show how this can be applied to estimate the states and likelihood for a two-dimensional Susceptible--Infected--Recovered (SIR) epidemic model \citep{Keeling:2007}, where we only observe a single component of the state. 
%This also estimates the filtering density, $p(x_t|y_t)$, if part of an SMC algorithm \cite{Doucet2001}.\\
In doing this, we begin to develop the idea that the importance sampling algorithm can be considered as a suitably modified version of the original process, conditioned on a set of event times that are initially randomly generated. 

Instead of defining the model in terms of the population numbers ($S$ and $I$) we instead work primarily in terms of the numbers of two events that can occur: infection and recovery \citep{JG12,BR14}. Thus we denote by $Z_1(t)$ and $Z_2(t)$ the number of infection and recovery events that have occurred up to time $t$, and hence the state of the system is specified by a vector $\mathbf{Z}(t) = (Z_1(t),Z_2(t))$.
We assume that our observations of the system correspond to the number of infection events, $y$, over the interval of time $[0,1]$, but recovery events are not observed; hence we do not know the number of infected or recovered individuals in the system. Defining $\mathbf{z}_i = \mathbf{Z}(i)$, the observation likelihood is 
\begin{equation}
	 p(y| \bz_1) = \delta_{g(\bz_1),\, y},
\end{equation}
where the function $g$ picks the first component of the vector $\bz_i$.\\

% \begin{figure}[tb]
% 	\centering
% 	%\includegraphics[]{}
% 	\caption{The SIR model. $Z_1$ and $Z_2$ count the numbers of infection and recovery events respectivley. }
% 	\label{fig:SIR_model}
% \end{figure}

First note that we can write the number of susceptible and infected individuals in terms of the number of events as (dropping the dependence on $t$),
\begin{equation}
	\begin{pmatrix}
	S\\
	I
	\end{pmatrix}
	=
	\begin{pmatrix}
	S_0\\
	0
	\end{pmatrix}
	+
	\begin{pmatrix}
	-1 & 0 \\
	1 & -1
	\end{pmatrix}
	\begin{pmatrix}
	Z_1\\
	Z_2
	\end{pmatrix}
	\label{eq:SIR}
\end{equation}
where the matrix in Eq \eqref{eq:SIR} is known as the stoichiometric matrix \citep{van_kampen}, and we have assumed we start with a completely susceptible population of size $S_0$. The stoichiometric matrix encodes how each event changes the numbers of $S$ and $I$. For example, an infection event decreases the number of $S$ by 1 and increases the number of $I$ by 1. The rates of the two events are,
\begin{equation}
	\begin{aligned}
	a_1 &= \beta SI = \beta (S_0-Z_1)(Z_1 - Z_2), \\
	a_2 &= \gamma I = \gamma (Z_1 - Z_2),
	\end{aligned}
	\label{eq:SIR_rates}
\end{equation}
where $\beta$ and $\gamma$ are the infection and recovery rate parameters respectively, and are herein assumed fixed.
The reason to work primarily in terms of the event counts rather than the population numbers is that event counts only ever increase, which leads naturally to the relation
\begin{equation}
	Z_2 \le Z_1.
	\label{eq:SIR_const}
\end{equation}

Given that the system starts in a particular state, $\bz_0$, at time $t=0$, a realisation can be generated using the SSA \citep{Gillespie:1976}. There are two basic versions of this; in the `direct' version, the time to the next event is exponentially distributed with rate parameter $a_0 = a_1+a_2$,
\begin{equation}
	t' \sim \text{Exp}(a_0).
	\nonumber
\end{equation}
The event that happens after this time (1 or 2) is then chosen randomly according to the probabilities $a_i/a_0$, $i=1,2$. An equivalent way of performing this simulation is to instead draw two times, $t_1'\sim \text{Exp}(a_1)$, and $t_2'\sim \text{Exp}(a_2)$, then the next event is chosen as the one with the smallest time. This is known as the next reaction method \citep{Gillespie:1976,Gibson2000,Anderson2007}. However generated, a realisation of the process can be specified by the initial condition, $\mathbf{Z}(0)$, and a list of times and the index of the event that occurs at those times,
\begin{equation}
	\psi = \left\{ \{ t_1, e_1\} , \{ t_2, e_2 \} ,\dots \{ t_n, e_n\} \right\}.
	\nonumber
\end{equation}
where $ 0 < t_i< 1$, and $e_i \in \{1,2 \}$ is the index of the $i$th event. 
The form of the rates of each event given in Eq.~\eqref{eq:SIR_rates} means that any realisation generated by this procedure will have Eq.~\eqref{eq:SIR_const} automatically satisfied. 
%But, if the probability of generating a realisation that matches our observations is low, then the Monte Carlo estimate of the marginal likelihood will have a large variance. 

We can use importance sampling, as introduced in the previous section, to generate realisations that match our data exactly. This means the number of infection events over the observation interval will be equal to $y$, but their times are random as are the number and times of recovery events. The basic idea is that a realisation can be produced by first randomly generating the times of $y$ infection events over the interval $[0,1]$ and then generating recovery times such that the set of all times are consistent with the model and the data. The recovery times are generated by simulating a modified version of the the model, conditioned on the initially-generated infection times. 
Consistency with the model means that Eq~\eqref{eq:SIR_const} must be true; consistency with the data means that as we specify the times of infection events in the first step, then these must always be possible, which requires 
\begin{equation}
	a_1>0 \implies I = (Z_1-Z_2)>0,	
	\nonumber
\end{equation}
otherwise the epidemic has faded out\footnote{This condition may not be true after the final observed infection event, depending on what other observations are made on the system afterwards.}.\\

To generate a realisation we proceed as follows. 
First, a set of $y$ ordered infection times is generated from a uniform distribution over the interval $[0,1]$,
\begin{equation}
	\tau = (t_1, t_2,\dots,t_y).
\end{equation}
We call these \emph{forced events}. 
Next we generate recovery times to produce a valid realisation; we do this by running a modified version of our model over the same interval of time, conditioned on the times $\tau$. To do this, first note that because we have specified the times, and the exact number, of the infection events in the first step, the rate of further infection events must be zero over the observation period. Also, if $I(t)=1$ then there cannot be another recovery event before the next infection event, hence this rate of the event must be zero. The modified process therefore has rates
\begin{equation}
	\begin{aligned}
	b_1 &= 0, \\
		b_2
		&=
		\begin{cases}
			a_2 \quad \text{if} \quad I>1,\\
			0\phantom{_2} \quad \text{if} \quad I=1,
		\end{cases}
	\end{aligned}
	\label{eq:SIR_modified_rates}
\end{equation}
for events of type 1 and 2 respectively and $b_0 = b_1+b_2$.
This may seem wasteful to specify the process in this way (with $b_1=0$), but this redundancy significantly simplifies the general version of the algorithm and its exposition. This can be removed---and the algorithm made slightly quicker---by simply re-labelling the events as discussed later.

Starting from $t=0$, the modified process is simulated with rates \eqref{eq:SIR_modified_rates}. Let $t_n$ be the time of the next forced event, given the current time $t$. The algorithm  then \emph{proposes} a time to the next event, drawn from an exponential distribution,
\begin{equation}
 	t' \sim \text{Exp}(b_0).
 	\nonumber
\end{equation} 
This is compared with $t_n$ as follows,
\begin{itemize}
	\item If $t' < t_n - t$ then the next event is a recovery event at time $t+t'$, that is, we set $Z_2 \leftarrow Z_2 +1$ and $t \leftarrow t + t'$.

	\item If $t' \ge t_n -t $, then the next event is an infection at time $t_n$, so $Z_1 \leftarrow Z_1 +1$ and $t \leftarrow t_n$, and the next forced event time, $t_n$, is updated. 
	%\rev{might introduce an index to count the number of these events}.
\end{itemize}
Note that if $b_2=0$ then the next step in the algorithm will be to implement the next forced infection event at time $t_n$. Once the algorithm has reached the time of the last forced infection event, then the procedure above carries on, potentially adding recovery times, but if $t+t'>1$, then this means that no new events occur before the end of the observation window and the simulation terminates. 

Thus at each iteration there are three things that can happen: either add a recovery at the proposed time, add a pre-calculated infection time, or the end of the observation window is reached. Through this procedure we generate a realisation where the number of infection events matches the observation, but is also consistent. What remains is to calculate the probability densities of the realisation under the modified process (from which it is generated) and under the original process, so as to calculate the weight of the realisation. 
First, the contribution to the weight from the initially-generated infection times is the same as before, given by Eq.~\eqref{eq:init_import}. The rest of the weight contributions are then most easily calculated iteratively as the algorithm proceeds.
First note that the log transition density for an event under the original process is 
	\begin{equation}
		\log\left(\frac{a_i}{a_0} a_0 \exp(- a_0 s)\right) = \log(a_i) - a_0 s, \quad i=1,2,
	\end{equation}
where $s$ is the time until the event. If a recovery event happens ($t' < t_n - t$) then the log importance weight is updated as 
	\begin{equation}
		w \leftarrow w + \log\left(\frac{a_2}{b_2}\right) - (a_0 - b_0)t'.
	\end{equation}
If an infection event occurs ($t' \ge t_n -t$) then the log importance weight is updated as,
	\begin{equation}
		w \leftarrow w + \log(a_1) - (a_0-b_0) (t_n-t), %- (-b_0 (t_n-t)),
	\end{equation}
where the second term is the probability of proposing a time greater than $t_n$. Finally, after the last infection event, if $t' > 1-t$ then no event occurs so the contribution is
\begin{equation}
	w \leftarrow w - (a_0-b_0) (1-t), % - (-b_0 (1-t))
	\label{eq:end_day}
\end{equation}
which is simply the log of the ratio of the probabilities of no further events in the interval under the original and modified processes.

\subsection{SEIR model}
\label{sec:SEIR}

We now extend the algorithm developed so far to generate realisations of an SEIR model that match observations exactly. This adds an additional complication in maintaining consistency of a particular realisation.
The SEIR model is similar to the SIR model, but has an additional latent class, $E$, where an individual is infected, but not yet infectious. This model can be defined in terms of three events: infection ($S\rightarrow E$), latent progression ($E\rightarrow I$) and recovery ($I\rightarrow R$). For conciseness, we will refer to these as events $1$, $2$ and $3$ respectively, and $\mathbf{Z}(t) = (Z_1(t), Z_2(t), Z_3(t))$ counts the number of these events that have occurred by time $t$. The relation between the number of each event and the population numbers is,
\begin{equation}
	\begin{pmatrix}
	S\\
	E\\
	I
	\end{pmatrix}
	=
	\begin{pmatrix}
	S_0\\
	0\\
	0
	\end{pmatrix}
	+
	\begin{pmatrix}
	-1 & 0 & 0\\
	1 & -1 & 0 \\
	0 & 1 & -1
	\end{pmatrix}
	\begin{pmatrix}
	Z_1\\
	Z_2\\
	Z_3
	\end{pmatrix},
\end{equation}
where we assume an initially susceptible population of size $S_0$.
The rates of each event are then
\begin{equation}
	\begin{aligned}
	a_1 &= \beta SI = \beta (S_0-Z_1)(Z_1 - Z_2),\\
	a_2 &= \sigma E = \sigma (Z_1 - Z_2),\\
	a_3 &= \gamma I = \gamma (Z_2 - Z_3),
	\end{aligned}
	\label{eq:SEIR_rates}
\end{equation}
where ($\beta,\gamma, \sigma$) are the fixed parameters and $a_0 = a_1+a_2+a_3$. For this model, we assume the number of event 2 ($E \rightarrow I$) is observed over the interval $[0,1]$, which is denoted $y$. This can be justified for diseases where the onset of symptoms and infectiousness coincide, such as influenza.

The simulation algorithm for this model is similar to that for the SIR model. In the first step, the times of event 2 are generated, then in the second step, the times of the other two events (1 and 3) are generated, conditional on the event 2 times. In the SIR model, we assumed that the observations were of the first event in the chain (infection). In this model we now observe the second event (when an individual becomes infectious after a latent period), which introduces additional complexity.
For this model, we now require for consistency that,
\begin{equation}
	Z_i \le Z_j \quad \text{for} \,\,\,i<j,
	\label{eq:seir1}
\end{equation}
as well as
\begin{equation}
	E+I=Z_1-Z_3> 0,	
	\label{eq:SEIR_model_const}
\end{equation}
which enforces that the disease cannot go extinct. In addition to these, because the event 2 times are specified in the first step, when they are implemented in the second step,
\begin{equation}
	a_2 > 0 \implies E = Z_1-Z_2 > 0.
\end{equation}
It should be emphasized that $E$ may go to zero (and this has to be allowed otherwise we change the dynamics of the original model), but if it does go to zero then an event 1 must occur before the next event 2. In practice, this means that at certain points in the simulation an event 1 may also need to be forced to maintain consistency of the realisation---how this is done is discussed later. Depending on what other observations have been made, we may also wish to condition the process on the final size of the outbreak. If $N_F$ is the total number of event 2 observed over the course of the outbreak then $Z_1 \le N_F$, which by Eq.~\eqref{eq:seir1} implies $Z_2 \le N_F$ also.

% Thus in this example we need to now force an event to ocure before the next $Z_2$ event. This is to maintain consistencey in the model, i.e. $Z_i \le Z_j$, $i<j$.

% Intuitivey this means that for an event $i$ to fire it must have $a_i>0$. If the state of the system is such that this is true then we need to force an event before the next event. Now instead of holding the times in a vector we store them in a stack. This is becasue we may need to add more times as the simulation proceeds. 

\begin{figure}[ht!]
	\centering
	\includegraphics[width=0.7\textwidth]{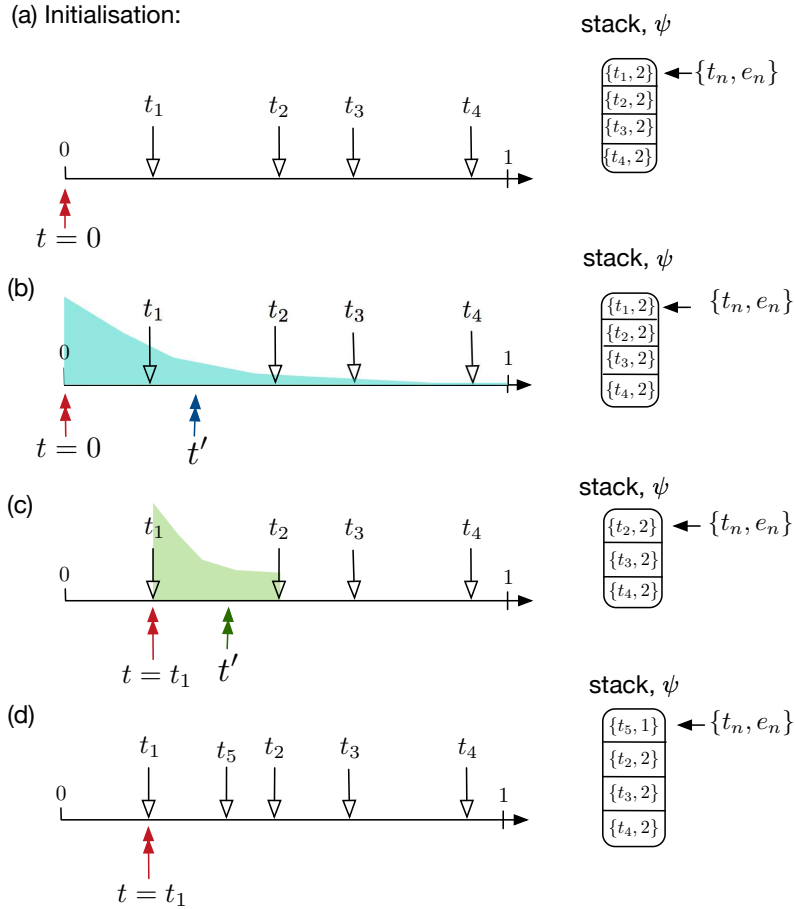}
	\caption{Illustration of the algorithm for the SEIR model, $\mathbf{Z}(0) = (2,1,0)$ and $y=4$.
	(a) For initialisation the times of the 2nd (observed) event are generated and added in reverse order to the stack, $\psi$; $t_n$ and $e_n$ point to the the top time and event index respectively.
	(b) In the first step, $E(0)=1$ and $e_n=2$ so the state of the system is consistent with the next forced event. Hence a time, $t'$, is proposed, with pdf shown in light blue. In this example, $t+t' > t_n$ so the next forced event is implemented, and popped off the top of the stack.
	(c) Now $E(t_1)=0$ and $e_n=2$ so the state is inconsistent. Thus the algorithm forces an event 1 in the interval $[t_1, t_2]$. The pdf for this time is shown in light green. (d) The stack after this addition, where $t_5=t_1+t'$. Note that after $t_5$ is added to the stack, $t=t_1$ still, but $e_n=1$ so the state is now consistent with the next forced event. Hence, the next step in the algorithm will be to propose a time for the next event as in (b).}
	\label{fig:SEIR_times}
\end{figure}

The simulation proceeds in two stages as follows. For the first stage, the $y$ times of event 2 are generated from a uniform distribution over the observation interval and sorted. Before, these were stored as a simple vector, but now these are stored in a stack denoted $\psi$ \citep{Knuth:2,AhoUllman:1995}.  These times are added to the stack in reverse order along with the event indices, so the earliest forced event is at the top of the stack. 
The move to a stack is because the algorithm may need to force more events during the course of the simulation.
Let $t_n$ and $e_n$ point to the top time and event index respectively, which we call the \emph{next forced event}. An example of this step is shown in Figure \ref{fig:SEIR_times}(a).

The second stage of the procedure simulates a modified Markov chain over the interval $[0,1]$ to generate the times of the other events. 
%An iteration of this begins by checking the index of the next forced event, $e_n$. If $e_n=2$ and $E=0$ then there must be event 1 within the interval $[t,t_n]$, or else the state would be inconsistent when the algorithm reaches the time of the next forced event.
The algorithm proceeds in a similar manner to the SIR model, by proposing times to the next event and accepting / rejecting that based on the time of the next forced event. The only additional step occurs at the beginning of each iteration, where the algorithm checks the index of the next forced event, $e_n$, and the state of the system.
This idea is illustrated by way of an example in Figure \ref{fig:SEIR_times}. At $t=0$ the state of the system is $\mathbf{Z}(0) = (2,1,0)$. $e_n=2$, but $E=1$, so the state is consistent with the next forced event. Thus a time, $t'$, is proposed derived from a modified process, specified in more detail later. In this example, $t+t' > t_n$ so the next forced event is implemented instead of the proposed one, and $t_n$ and $e_n$ are updated. After this event, $t=t_1$ and $\mathbf{Z}(t)=(2,2,0)$. For the next iteration, $e_n=2$ but $E=0$, hence the state is inconsistent with the next forced event so at least 1 event 1 must occur within the interval $[t_1,t_2]$ for the model to be consistent at time $t_2$. 
The time of this event is generated from a truncated exponential distribution on this interval, with rate $a_1(t)$, i.e., the current rate of this event,
\begin{equation}
	t' \sim \text{TruncExp}(a_1,0,t_n-t).
\end{equation}
The time $t+t'$ and event index 1 are then pushed on to the top of the stack. This step is represented pictorially in Figure \ref{fig:SEIR_times} (c and d).
%This event is taken to be the \emph{first} time a type 1 event occurs within that interval.
Whenever an additional event is forced in this way, the modified rate of that event ($b_1$ in this case) is set to zero until the event has been implemented.
The contribution to the log importance weight of this extra forced event is
\begin{equation}
	w \leftarrow w -\log(a_1) + a_1 t' +\log[1-\exp(-a_1 (t_n-t))],
\end{equation}
which is the log of the density of a truncated exponential RV with rate $a_1$. 
%By specifying this as the first event within the interval, the modified rate of that event, $b_1$, must be 0, up until the event occurs. So more events of this type can occur after, but not before. 
This step to ensure consistency is different from the scheme presented in \citet{McKinley:2014}. The consequences of this are discussed fully later.

After this step, the algorithm proceeds essentially as for the SIR model with a few modifications. First the modified rates are calculated as follows:
\begin{equation}
	\begin{aligned}
	b_1 &= 
		\begin{cases}
			0\phantom{_1}   \quad \text{if} \quad e_n = 1 \quad \text{or} \quad Z_1 = N_F,\\
			a_1 \quad \text{otherwise},
		\end{cases}\\
	b_2 &= 0, \\
	b_3 &=
		\begin{cases}
			a_3 \quad \text{if} \quad I+E>1,\\
			0\phantom{_2} \quad \text{if} \quad I+E=1,
		\end{cases}\\	
	\end{aligned}
	\label{eq:SEIR_modified_rates}
\end{equation}
and $b_0 = b_1+b_2+b_3$. The second rate is always zero as the number and times of these events are already generated. The third rate is set to zero if there is only a single infected or exposed individual, to stop the disease going extinct. The first rate is set equal to zero if $e_n=1$, i.e., if event 1 is the next forced event. The first rate is also set to zero if the number of these events equals the observed final size (if this is available). Thus all realisations would end with the correct total number of infections. We can see that there are situations in which $b_0 = 0$, in which case the next forced event is implemented at the next step in the algorithm. \\

Once the modified rates are calculated, a time is proposed, $t' \sim \text{Exp}(b_0)$ and compared to the time of the next forced event, $t_n$ as described for the SIR model. The only additional step required is that if the proposed time is accepted, the particular event that occurs is randomly chosen in proportion to its rate $b_i/b_0$, $i \in \{1,3\}$ as in the SSA. Hence if the proposed time is accepted ($t'< t_n -t$) and the $i$th event is chosen then the log importance weight is updated as
	\begin{equation}
		w \leftarrow w + \log\left(\frac{a_i}{b_i}\right) - (a_0 - b_0)t'.
	\end{equation}
If instead, the next forced event is implemented, ($t' \ge t_n -t$) and $e_n$ is the index of that event, then
	\begin{equation}
		w \leftarrow w + \log(a_{n_e}) - (a_0 - b_0) (t_n-t).
	\end{equation}
After these steps, the algorithm returns to the start and checks the state of the system and the next forced event. This continues until after the last detection event, at which point the constraints on the modified process are much simpler as the algorithm only has to stop fade out of the disease and there are no more forced events. Once the first proposed time goes beyond the observation window ($t+t'>1$) then the algorithm stops and the final weight contribution of this step is as in Eq.~\eqref{eq:end_day}.

The importance sampling described here does have one limitation, which is that it will not perform well for all parameters. 
For example, if $1/ \sigma\rightarrow 0$ then the model effectively becomes an SIR model as an individual leaves the exposed class almost immediately after entering it. In this case the times of the infection events and the observed events become highly correlated and the importance sampling process is no longer a good approximation to the true process. In this case the the importance sampling can actually increase the variance of the estimate of the likelihood in comparison to a more naive approach.
This does restrict the range of parameters for which the algorithm can be used. If used within a pmMH routine, then this can be enforced by setting appropriate priors for the Metropolis Hastings part of the algorithm.

\subsection{Summary of the exact-matching algorithm}
\label{sec:algo}

At this point we can give a summary of the importance sampling algorithm. Assume we have a model with $M$ event types, where we observe the number of type $k$ over the interval $[0,1]$. The algorithm is then as follows, where steps 2 and 3 are model dependent and hence are discussed in more detail in the next two sections.

\noindent
\emph{Initialisation}: set the initial condition, $\mathbf{Z}(0)$, and generate the times of the $y$ observed events (of type k) from a uniform distribution over the interval $[0,1]$. Sort and add these to the stack, $\psi$, in reverse order and set $e_n$, $t_n$ and $t=0$.\\
Set the initial importance weight, $w=\ln(y!)$.
\begin{enumerate}

	\item Calculate the rates of the original process, $a_i(\mathbf{Z}(t))$, $i=1,\dots,M$ given the current state and $a_0 = \sum_{i=1}^M a_i$.\label{start}\\

	\item Check the consistency of system given the next forced event, $e_n$. If inconsistent, then force an event to fix this within the interval $[t, t_n]$. Assuming this event is of type $l$, generate 
	\begin{equation}
		s \sim \text{TruncExp}(a_l,0,t_n-t),
		\nonumber
	\end{equation}
	and
	\begin{equation}
		w \leftarrow w -\log(a_l) + a_l s +\log[1-\exp(-a_l (t_n-t))].
		\nonumber
	\end{equation}
	Push $t+s$ and event index $l$ onto the stack and update $e_n$ and $t_n$.\\

	\item Calculate the rates of the modified process, $b_i(\mathbf{Z}(t), n_e)$, $i=1,\dots,M$, which depend on the current state and the next forced event. Calculate the total rate, $b_0 = \sum_{i=1}^M b_i$.\\

	\item Propose a time to the next event, $t'\sim \text{Exp}(b_0)$.\\

	\item If $t'< t_n-t$, choose an event index, $j\in \{1,\dots,M\}$ with probability Pr$(j=i) = b_i /b_0$ and 
	update $Z_j \leftarrow Z_j + 1$, $t\leftarrow t+t'$ and
	$$
	w \leftarrow w + \log\left(\frac{a_j}{b_j}\right) - (a_0 \rev{-} b_0)t'.
	$$

	Otherwise (if $t' > t_n-t$) implement the next forced event at time $t_n$.\\
	Set $Z_{n_e} \leftarrow Z_{n_e} + 1$, $t\leftarrow t_n$ and
	$$
		w \leftarrow w + \log(a_{\rev{e_n}}) - (a_0 - b_0) (t_n-t).
	$$
	\item While $|\psi| > 0$, goto \ref{start}.
\end{enumerate}
Once all the forced events have been implemented (the stack is empty, $|\psi|=0$), then the simulation continues until $t+t'>1$, but step 2 is no longer required and the conditioning of the modified process in step 3 is in turn simpler. The algorithm terminates once $t' > 1-t$ and the final contribution to the weight is
\begin{equation}
	w \leftarrow w - (a_0-b_0) (1-t).
	\nonumber
\end{equation}

As signposted, steps 2 and 3 depend on the model structure. For the the SIR model, step 2 is not required as we observe the first event in the chain. For the the SEIR model, these steps are detailed in Section \ref{sec:SEIR}. In the next Section we will describe how these steps can be carried out for a more complex model that potentially requires forcing chains of events to maintain consistency of the state given the next forced event.

\section{Model with symptomatic phases}
\label{sec:asym_model}

\begin{figure}[th]
	\centering
	\includegraphics[width=0.5\textwidth]{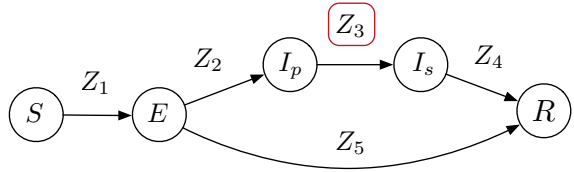}
	\caption{Model where individuals go through a pre-symptomatic stage and also includes asymptomatic individuals that do not contribute to the force of infection. Individuals are observed when they first enter the $I_s$ class.}
	\label{fig:big_small}
\end{figure}

\rev{In this section we give an example of how the algorithm can be applied to a model with a more complex structure that requires a decision tree for deciding how to force events to maintain consistency of a realisation. In the second part we incorporate this into a particle filter and compare it with a standard approach from the literature. The model has} the structure illustrated in Figure \ref{fig:big_small}. This splits the infectious class into two stages modelling a pre-symptomatic ($I_p$) and a symptomatic ($I_s$) phase \citep{Regan2016}. It also includes the possibility of asymptomatic individuals that do not contribute towards the overall force of infection (and hence go straight to the $R$ class). In this paper we refer to this as the SEIAR model. We assume that we observe $y$ of event 3 over interval the interval $[0,1]$, which corresponds to individuals becoming symptomatic. We also assume a final size observation, $N_F$, had been made from later observations. Final size here refers to the total number of detections over the course of the outbreak, not the total number of individuals infected.

The rates of the events for this model are:
\begin{equation}
	\begin{aligned}
		a_1 &= (S_0-Z_1)[\beta_p (Z_2-Z_3)+\beta_s (Z_3-Z_4)],\\
		a_2 &= q\sigma (Z_1-Z_2-Z_5),\\
		a_3 &= \gamma (Z_2-Z_3), \\
		a_4 &= \gamma (Z_3-Z_4),\\
		a_5 &= (1-q) \sigma (Z_1-Z_2-Z_5), \\
	\end{aligned}
	\label{eq:BFM_rates}
\end{equation}
where $(\beta_s,\beta_p,\sigma,\gamma,q)$ are fixed parameters, with $\beta_p$ and $\beta_s$ the transmission rates of pre-symptomatic and symptomatic individuals respectively.  

Only steps 2 and 3 of the algorithm are different to what has already been presented. Step 2 is more complicated than for the previous two models as there can now be situations where the algorithm needs to force a \emph{chain} of extra events so that the state is consistent at the time of the current next forced event.
For example, if $e_n=3$ and both $I_p=0$ and $E=0$ then at least 1 of each of event 1 and event 2 must occur in the interval $[t,t_n]$. Another complication arises in step 3 as we also need to monitor the number of event 5 (non-detections) that can occur, so that at later time-steps---as would occur in a particle filter---there are still enough susceptible individuals left such that the algorithm can match the final size observation.

There are different ways of implementing step 2, but we adopt the rule that when the current state of the system is inconsistent with the next forced event, we force the first event (in a possible chain of events) such that the system is again consistent. Returning to the example above, 
%\rev{May need a figure for this idea.}
if $e_n=3$ and both $I_p=0$ and $E=0$, the algorithm would first force an event 1 in the interval $[t, t_n]$, after which $n_e=1$, which is consistent with the current state. Once the algorithm has then reached this time, $E=1$, $I_p=0$ and $e_n=3$, so the state is again inconsistent and this is fixed by forcing an event 2 (which is now allowed because $E>0$).

This rule is adopted because the state of the system can be made consistent by forcing a single event, thus the algorithm only needs to keep track of the next forced event rather than keeping track of chains. Following on from this, the times of these extra forced events and the importance weights remain simple to calculate as detailed in the previous section 
and the algorithm proceeds through steps 3-5 with no further modifications. 
This rule also means that the logic of which events have to be forced, given the type of the next forced event, $e_n$, and the current state, $Z(t)$, can be represented as a decision tree that can be readily deduced.

% \begin{figure}[tb]
% 	\centering
% 	\includegraphics[width=0.33\textwidth]{fig4.eps}
% 	\caption{Decision tree for step 2 of the algorithm to simulate the SEEIIR model. The tree is traversed until it either forces an event within the interval $[t,t_n]$, or it reaches a `$\phi$' which indicates that the current state is consistent with the next forced event so no extra event needs to be forced.}
% 	\label{fig:dtree}
% \end{figure}

\begin{figure}[th]
	\centering
	\includegraphics[width=0.4\textwidth]{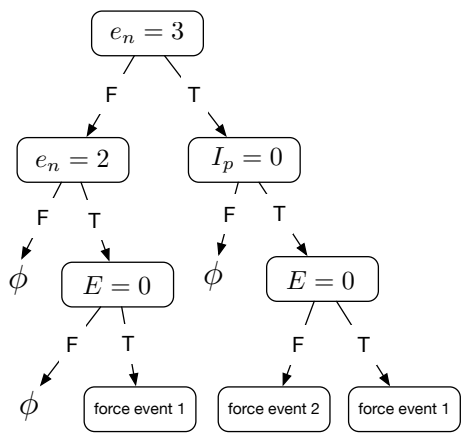}
	\caption{Decision tree for the SEIAR model with pre- and a-symptomatic individuals. If a `$\phi$' is reached, this indicates that the state, $\mathbf{Z}(t)$ is consistent with the next forced event.}
	\label{fig:dtree}
\end{figure}

The decision tree for computing the type, $l$, of the forced event in step 2 of the algorithm for this model is shown in Figure \ref{fig:dtree}. This is implemented as a set of conditional statements, where if a `$\phi$' is reached this indicates that the current state of the system is consistent with the next forced event (hence nothing extra needs to be forced). Once the type is computed, this index along with a time is added to the stack as described in Section \ref{sec:algo}.
Another example, with a simpler decision tree, is given in the Supplementary material. 
% For the present model, 
% This is needed as in this model an event 2 may be forced (and hence added to the stack), but within the interval of time until it occurs, an event 5 could also occur, the result of which may be that $E=0$. Hence after this event, the state is inconsistent which should be fixed by forcing an event 1.
After this step, the modified rates for this model are then calculated as follows:
% For example, if $e_n=3$ and both $I_p=0$ and $E=0$, we again require a chain of events to happen. We put the first one in, which then allows the second one to be inserted. But after this, it is possible a type 5 occurs, which then means $
%E=0$ again, but the next forced event is a type 2. Thus the state is again inconsistent. This is fixed by forcing another type 1 event. This is illustrated in figure ?? \\
%The decision tree is shown in Figure \ref{fig:dec_tree1}, and the modified rates are:
\begin{equation}
	\begin{aligned}
	b_1 &= 
		\begin{cases}
			0\phantom{_1}   \quad \text{if} \quad e_n = 1 \quad \text{or} \quad (Z_1 = N_F \,\, \text{and} \,\, q=1),\\
			a_1 \quad \text{otherwise}.
		\end{cases}\\
	b_2 &= 
		\begin{cases}
			0\phantom{_2} \quad \text{if} \quad e_n=2 \quad \text{or} \quad Z_2 = N_F,\\
			a_2 \quad \text{otherwise},
		\end{cases}\\	
	b_3 & = 0\\
	b_4 & = 
		\begin{cases}
			0\phantom{_2} \quad \text{if} \quad Z_1-Z_4-Z_5 = 1\\
			a_4 \quad \text{otherwise},
		\end{cases}\\		
	b_5 &=
		\begin{cases}
			0\phantom{_5}  \quad \text{if} \quad Z_5=S_0-N_F \quad \text{or} \quad 
			Z_1-Z_4-Z_5 = 1,\\
			a_5 \quad \text{otherwise}.
		\end{cases}\\	
	\end{aligned}
	\label{eq:BFM_mod_rates}
\end{equation}
The rates $b_1$ and $b_2$ are set to zero if either of these events have already been forced and are in the stack. The rates $b_1$ and $b_5$ are modified so as not to allow too many of either of these events when final size data is available. Finally, the rates $b_4$ and $b_5$ are also modified so that the disease cannot prematurely fadeout. Once the modified rates are calculated, all other steps of the algorithm are the same as Section \ref{sec:algo}.

\subsection{Inference example}
\label{sec:example}

In this section we use the SEIAR model to perform inference on an example time series using the pmMH algorithm. The particle filter uses importance sampling as described in the previous sections and the resulting posteriors are compared with those obtained by using an alive particle filter \citep{DelMoral:2015,Drovandi:2016}. The alive filter works at each time step by simulating a number of realisations using the SSA, until a certain number of matches  are obtained. This is clearly computationally expensive, but gives an unbiased estimate of the likelihood. 

For the SEIAR model we can define $R_0$, the basic reproductive ratio, as
\rev{\begin{equation}
	R_0 = \frac{q(\beta_p+\beta_s)}{\gamma},
\end{equation}}
where we have assumed frequency-dependent transmission and hence $\beta_s$ and $\beta_p$ are both scaled by $S_0-1$. We define $\kappa$ as the proportion of transmission due to pre-symptomatic individuals, (in class $I_p$). Hence the proportions of $R_0$ attributable to pre- and symptomatic individuals are \rev{$R_0^p = q\beta_p/\gamma = \kappa R_0$} and \rev{$R_0^s = q\beta_s / \gamma =   (1-\kappa)  R_0 $}, respectively. 

\rev{We performed inference on a number of synthetically generated outbreak time series from increasingly large populations. The parameters used to generate the data were set as $R_0=2.2$, $\kappa=0.7$, $1/\sigma =1$, $1/\gamma = 1$, $q=0.9$, which are similar to influenza, in populations of $N=150, 350, 500$ and $1000$. These could represent, for example, outbreaks aboard ships. A fixed initial condition of $Z(0) = (1,1,0,0,0)$ was used for simplicity, and this was replicated in the inference routines. Only major outbreaks were chosen and the final number of detections for each outbreak, $N_F$, are summarised in Table \ref{tab:tab1}. The time series themselves are plotted in the supplementary material.}

Both algorithms were coded in C. The Metropolis Hastings part of the algorithm used a simple random walk proposal and a pilot run was carried out to determine an appropriate covariance matrix for this, \rev{which was used for all four time series.} 
In our implementation of the alive filter we follow \citet{Drovandi:2016} and set a maximum number of trials at each time step, $K$, before the filter terminates and returns zero for the likelihood. This introduces some error to the algorithm, but stops it becoming stuck if the proposed parameters mean the observation is a very rare event; we set $K=10^5$ \rev{for the three smaller datasets, but this had to be increased to $10^6$ for the largest ($N=1000$). Using the smaller value for this resulted in a large error in the tails of the posterior for the parameter $\kappa$, as the likelihood is too small to estimate with only $10^5$ trials at each time step.} 

The alive filter was coded so that an iteration terminates as soon as it becomes inconsistent with the observation and the other constraints, which results in the optimal performance. For example, if a realisation matches the observation, but $E+I_s+I_p=0$ (so the disease has faded out) then the weight of the realisation was set to 0. Thus the estimate of the likelihood at each time step are (on average) the same from both filters. \rev{The particle filter using importance sampling, re-sampled the particles after each time step; no advantage was found by re-sampling less frequently.
We assume informative priors on both $1/\gamma$ and $1/\sigma$, of Gam$(10,1/10)$, with lower bounds of $0.5$ and $0.1$ respectively. This is done as without extra information, either from more data or other observations, these parameters are unidentifiable. For the other parameters, $R_0$,  $q$ and $\kappa$, we assume uninformative priors of U$(0.1,8)$, U$(0.5,1)$, U$(0,1)$ respectively.}

\begin{figure}[ht]
	\centering
	\includegraphics[width=0.8\textwidth]{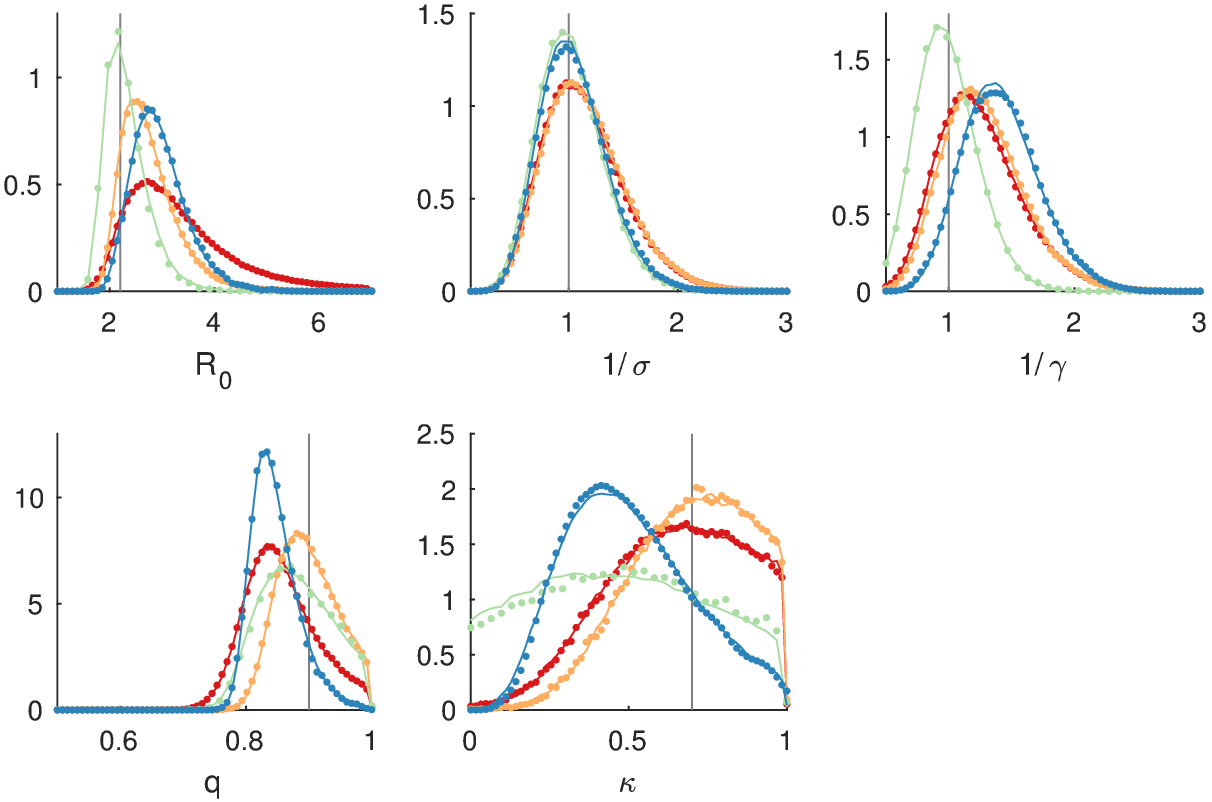}
	\caption{Marginal posterior distributions from performing inference using a particle filter with importance sampling (solid line) and using the alive filter (dots). $N=150$ (red), $N=350$ (yellow), $N=500$ (green) and $N=1000$ (blue). The true values of the parameters used to generate the data are marked by the grey lines.}
	\label{fig:results}
\end{figure}

\rev{The resulting marginal posterior distributions are shown in Figure \ref{fig:results} and effective sample sizes (ESS) per second of CPU time are given in Table \ref{tab:tab1}. Complete run times and statistics are given in the Supplementary material. Firstly we see that there is excellent agreement between the alive filter and importance sampling. As the data sets grow in size, both versions slow down due to increasing number of particles being used and the increasing number of events that have to be simulated. The speed up of the importance sampling filter over the alive filter, quantified in terms of the ESS per second of CPU time, increases as the size of the datasets also increases.} \rev{No attempt was made to tune the number of particles for either algorithm beyond attaining reasonable performance, so it is likely slightly better results could be obtained for both. The slight deviation between the posteriors for the parameter $\kappa$ for the $N=500$ and 1000 datasets is the result of the error introduced by too small a value of $K$ in the alive filter (see above).}

\begin{table}[ht]
\centering
\rev{\begin{tabular}{c|c|c|c|c}
$N$ & $N_F$ & particles & ESS s$^{-1}$ & speed-up \\
\hline
150 & 121 & 20 & 4.1 & 8.5 \\
350 & 288 & 40 & 2.0 &  10 \\
500 & 383 & 60 &  0.77 & 18 \\
1000 & 790 & 100 & 0.30 & 21 \\
\end{tabular}}
\caption{\rev{Inference statistics: $N$ is the population size, $N_F$ the total number of detected cases. The effective sample size per second of computing time (for the parameter $q$) is given along with the speed up over the alive filter. The ESS s$^{-1}$ for the other parameters are slightly different but follow the same pattern. All raw ESS values and running times are given in the Supplementary material.}
% Note that for $N=1000$, the alive filter was run with 100 particles and the version using importance sampling 200.
}
\label{tab:tab1}
\end{table}

\section{Discussion}
\label{sec:disc}

We have shown how importance sampling can be used to produce weighted realisations of a model that exactly matches data on the number of a given event over some interval of time. We can visualise the basic SSA and the exact matching algorithm as two ends of a continuous spectrum of potential importance sampling schemes. The SSA is essentially blind, sampling the transition density without regard to the observed data---a simple form of rejection sampling. Conversely, the exact-matching algorithm is guided at each state to ensure that the realisation is consistent with all the observations.
The importance sampling can be used to construct a particle filter for use in pmMH, providing a large speed-up in terms of ESS per unit time compared to bootstrap sampling using the alive filter.

The importance sampling builds on the work of \citet{McKinley:2014}, but differs in two major respects.  Firstly, the algorithms in \citeauthor{McKinley:2014} are tailored explicitly to the SIR and SEIR models where recovery events are observed. \rev{In contrast, the algorithm presented here (summarised in Section \ref{sec:algo}) is general in that it can be applied to any continuous-time Markov chain where a single event is observed. The other key difference is how the current algorithm forces extra events to maintain the consistency of a realisation. In the older algorithm, when it was detected that an extra event had to be forced, and there is more than one event possible, then the algorithm chooses the event in proportion to the relative rates of the two. This implicitly incorporates the consistency requirements but
is prone to numerical instabilities as many events can occur in the interval, in effect reducing the size of the interval until the next forced event, but without actually making the state consistent. Hence the algorithm attempts to correct for this by putting events into smaller and smaller time intervals, where at some point errors in floating point arithmetic can arise.} 
\rev{In contrast, the current algorithm explicitly forces particular events, in a specific order, (the type of which is encoded by the decision tree) and does not automatically update the current time to these as the original algorithm does. This has two effects; firstly, other events can be simulated in the intervening time periods, which may reduce the need to force further events later on. Secondly, consistency is ensured without undue forcing and is less likely to produce realisations that deviate from expected behaviour that would then be assigned a low weight. The knock on effect of this that numerical instabilities are reduced, as the need to force events into very small intervals is reduced.}

\rev{Still, numerical instabilities can arise in this algorithm. Generating truncated exponential random variables on small intervals has inherent instabilities due to the exponentiating required in calculating the cdf. This source of error could be circumvented by instead generating times from a uniform distribution when the size of the interval is below some threshold value (also taking into account the different contribution to the weight). Another workaround is to simply set the weight of realisations that become inconsistent to zero. Such an egregious realisation is likely to have a very small weight anyway, so this is unlikely to result in any error overall (as long as some remaining particles have a positive weight).}

\rev{Another standard approach for epidemic inference problems is data-augmented MCMC \citep{Gibson:1998,ONeill:1999}. In contrast to pmMH, which marginalises over this missing data in the calculation of the likelihood, data-augmented MCMC infers the missing data (the exact times of the events) as part of the overall Markov chain. When all event times are known, the likelihood is trivial to write down and conditional distributions for the parameters can often be derived, allowing for efficient Gibbs sampling. The greatest strength of the data-augmented approach is its flexibility; non-Markov models are handled as easily as Markov ones along with potentially large amounts of heterogeneity in the population and spreading process \citep{Jewell:2009,Cowling:2015,Stockdale:2017}.
The downside is that there is strong dependence between the missing data and the parameters that means mixing can become very slow and convergence can become an issue as the amount of missing data increases \citep{McKinley:2014,Pooley_2015,Walker:2017}.}

\rev{In contrast, particle filters marginalises over the missing data in estimating the likelihood. This means that the MCMC scheme targeting the parameter posterior is much simpler and easy to tune.}
\rev{Another important difference between data-augmented MCMC and pmMH is how they deal with increasing amounts of data, collected from independent outbreaks. One aspect of data-augmented MCMC is that it is essentially a serial algorithm. Thus performing inference over many independent outbreaks becomes challenging for the same reasons mentioned above; the state space becomes so large that convergence becomes a problem. Parallel chains can be run, but if convergence is an issue, this is not that useful.  In contrast, because they marginalise over missing data, particle filters can easily be parallelised and hence take advantage of modern computing hardware. 
For the particle filter this is most easily accomplished by running a number of independent filters on separate CPU cores and averaging the the results to obtain an estimate of the likelihood with lower variance \citep{Drovandi:2014}. The SMC$^2$ approach described below takes this parallelism even further as all parameter particles at a particular iteration can be updated independently. }

 The main drawback of pmMH is that the mixing of the main chain depends strongly on the variance of the likelihood estimate. 
There is a trade off between decreasing the variance of the log-likelihood and increasing the number of particles and hence the computational expense. Thus a higher variance estimate resulting in worse mixing can be offset by reduced computational expense and hence more samples from the posterior. For idealised models, the optimal performance is achieved by tuning the variance to be in a particular range \citep{Pitt:2012,Doucet2015,Sherlock2015}, which in turn maximises the effective sample size (ESS) per unit of computational time \citep{Sherlock2015}. In this paper we have not attempted to tune either of the particle filters employed herein, so it is likely better performance can be obtained for both. Recently, SMC$^2$ algorithms have been proposed to ameliorate these tuning issues \citep{Drovandi:2016,Golightly:2017}. These use sequential Monte Carlo to target both the parameter posterior as well as the states of the system. The importance sampling developed in this paper can be easily used in these approaches.

% Thus we adopt this same measure of performance and compare the \rev{maximum/minimum?} ESS per unit time achieved when performing inference using a pmMH algorithm.

%In this paper we have chosen to present this algorithm through a series of examples rather than in full generality. One reason for presenting examples is that the details matter, and I have tried to emphasise this. This also touches 

The primary weakness of \rev{the importance sampling presented in this paper} is that although the algorithm is general, it is not a black box. The implementation depends on the model structure, the rates of the events, as well as the event observed and other constraints. Details and edge cases are important for the algorithm to be correct, and testing is not always simple as we are working with rare events. As an example of an edge case, if calculating the likelihood for the last point in a time series, the simulation will depend on whether the disease is allowed to fade out after all forced events have been implemented. It should be clear that models where we observe the first event in a longer chain are much easier to handle than those where we observe later events and multiple events may need to be forced to maintain consistency. This is unavoidable when using a forward simulation approach.

Another weakness of the importance sampling scheme is that it breaks down if the rate of the observed event become too large. For example in the SEIR model, if $1/ \sigma\rightarrow 0$ then the model effectively becomes an SIR model as an individual leaves the exposed class almost immediately after entering it. In this case the times of the infection events and the observed events become highly correlated and the importance sampling process is no longer a good approximation to the true process. This limitation also applies to the algorithms presented in \citet{McKinley:2014}.
This problem essentially stems from attempting to do model selection (between an SIR and SEIR model) at the same time as parameter inference. To avoid this when running the pmMH algorithm we simply set priors that disallow such large rates, meaning that the SEIR model cannot become an SIR model. In the example in Section \ref{sec:example}, there was a lower bound on the parameter $1/\gamma$ of 0.5, i.e., the latent period must be at least half a day. It is natural that there be a trade-off between the speed of a method and generality of the problems that it can be applied to. The alive filter does not have any restrictions on the parameters, but is much slower.

We have provided MATLAB code for all of the models presented in this paper \citep{Epistruct}. This code is somewhat unoptimised, to keep it simple to follow. 
The particle filters used in the inference example (Section \ref{sec:example}) were coded in C as this gives an order of magnitude improvement in speed.
There is some redundancy in how we have specified the algorithms. For example, the modified rates of the observed event are always zero. This was done for readability and is easily factored out for some performance gains. This can be done by simply relabelling the events such that the observed event is the last. For example, in the SEIAR model, we would relabel 
\begin{equation*}
	(1,2,3,4,5) \rightarrow (1,2,5,3,4). 
\end{equation*}
Then, in the vector of propensities, $(a_i)$, the event with rate zero is always as the end and hence is never iterated over when generating the next event type.
It is also possible to re-factor the algorithm to remove the need for a stack to hold the forced event times, which would allow further optimisation when in specific states. Such optimisation comes at the expense of generality and clarity, so we have not presented this here.

%There are other ways the modified simulations could be carried out. In adding chains of events one could works backwards. The problem here is that it is not obvious what distribution the times of these should be drawn from. In working forward a truncated exponential is natural approximation, as we cannot calculate the distribution simple computationally efficient manner. Another possibility is to specify that there must be at least 1 of a type of event within an interval (and the time can be inserted). The problem with this is that Markov chain is then must be conditioned on these future events and calculating the transition density for such as process is again a hard problem. The algorithm does not exploit the times that are initially precomputed. For example is two detection events happen close together, then it is more likely that the events earlier in the chain happen before both of these, and this could be promoted by altering the parameters (being careful not to set any rates to zero that are >0 in the original process.) Again this could improve the algorithm but at further computational expense. 

The importance sampling algorithm described in this paper is state dependent, but we do not alter the underlying parameters of the model. Other schemes for rare event simulation \citep{Roh:2010} are based on altering the parameters of the process to create more matches and the use of cross-entropy methods to guide this. Similar ideas could be implemented here, but the computational expense would probably outweigh the benefit. 
The exact-matching algorithm is easily extended to the situations where the observations are noisy rather than exact. Instead, some number of events is sampled from the observation density (consistent with the observation) and then the simulation algorithm is run for that value.  Noisy observations increase the performance of a bootstrap particle filter using the SSA for sampling because it allows a particle to match a larger set of states.
No gain would be seen using the algorithm presented in this paper, as the number of observed events is set exactly for each realisation generated. Note that the SEEIIR model described in \rev{the Supplementary material} could be written using a binomial observation process, instead or observed and unobserved events, but this would still benefit from using importance sampling to produce realisations more likely to match the observations.

In this paper we have assumed observation of a single event type, which is natural when modelling epidemics, but not necessarily for other systems. In ecology, the Lokta-Volterra model often assumes there are observations of two population numbers (predators and prey). In this case there are a range of numbers of events that could give rise to an observation. 
In principle, we can use similar ideas to those presented here to construct realisations that match these types of observations. The difficulty arises in generating the times of the forced events such that they are ordered correctly, and then also calculating the order statistics. Such approaches are currently under investigation.

%In summary, we have described an importance sampling algorithm that can produce realisations of a Markovian model that exactly match observations. When used in a pmMH algorithm, we obtain a large speed up in the efficiency compared with using simple bootstrap filtering. 

\begin{acknowledgements}

This research was supported by an ARC DECRA fellowship (DE160100690). AJB also acknowledges support from both the ARC Centre of Excellence for Mathematical and Statistical Frontiers (CoE ACEMS), and the Australian Government NHMRC Centre for Research Excellence in Policy Relevant Infectious diseases Simulation and Mathematical Modelling (CRE PRISM$^2$).
Supercomputing resources were provided by the Phoenix HPC service at the University of Adelaide. 
AJB would also like to thank Joshua Ross and James Walker for comments on an earlier draft of the manuscript.
\end{acknowledgements}

% \bibliographystyle{spcustom}
% \bibliography{adelaide_refs.bib}

% %

\end{document}